\documentclass[9pt,sigconf]{acmart}

\usepackage{booktabs} 

\graphicspath{{figure/}{figures/}}

\setcopyright{rightsretained}

\acmDOI{10.475/123_4}

\acmISBN{123-4567-24-567/08/06}

\acmConference[NetAI'19]{ACM SIGCOMM 2019 Workshop on NetAI}{August 23, 2019}{Beijing, China} 
\acmYear{2019}
\copyrightyear{2019}

\acmPrice{15.00}

\begin{document}
\title{Smart Prediction of the Complaint Hotspot Problem in Mobile Network}

\author{Lin Zhu, Juan Zhao, Yiting Wang, Juanlan Feng, Chao Deng, Hui Li}
\orcid{1234-5678-9012}
\affiliation{%
	\institution{China Mobile Research Institute}
	\streetaddress{Address}
	\city{Beijing} 
	\state{China} 
	\postcode{Zipcode}
}
\email{{zhulinyj, zhaojuan, wangyitingyjy, fengjunlan, dengchao, lihuidsj}@chinamobile.com}

\renewcommand{\shortauthors}{L. Zhu et al.}

\begin{abstract}
In mobile network, a complaint hotspot problem often affects even thousands of users' service and leads to significant economic losses and bulk complaints. In this paper, we propose an approach to predict a customer complaint based on real-time user signalling data. Through analyzing the network and user sevice procedure, 30 key data fields related to user experience have been extracted in XDR data collected from the S1 interface. Furthermore, we augment these basic features with derived features for user experience evaluation, such as one-hot features, statistical features and differential features. Considering the problems of unbalanced data, we use LightGBM as our prediction model. LightGBM has strong generalization ability and was designed to handle unbalanced data. Experiments we conducted prove the effectiveness and efficiency of this proposal. This approach has been deployed for daily routine to locate the hot complaint problem scope as well as to report affected users and area.

\end{abstract}

%
%




\maketitle

\section{Introduction}
The structure of the communication network is complex with the huge number of network elements, which brings great challenges to daily operation and maintenance management. The end users usually make complaints to the customer service when their service has been affected. When a type of business complaints exceeds the threshold, a hot spot for complaints will be established by the customer service. It is necessary to increase the processing efficiency as much as possible and reduce the scope of influence. Usually, the network department will locate the problem of network through the analysis of the health of network elements, cells and the signalling data with expert experience after receiving the hotspot complaints. However, the procedure is not much efficient and cannot monitor the complaint hotspot problem at first due to the post-complaint processing. Therefore, there is an urgent requirement for automatic monitoring method to solve the complaint hotspot problem.

Recently, some scholars and experts have carried out research on complaint analysis and problems location based on user complaint information or equipment KPI indicators. There are three difficulties in the prediction of complaint hotspot problem: 1) Due to the inaccuracy of the customer complaints, the detection method based on raw complaint information may be inaccurate; 2) The customer service receives the complaints after users experience is affected for a period of time, that is customer complaints tend to be later than network problem. Therefore, the problem location method based on the complaint information cannot locate the complaint hotspot problem at first; 3) The complaint hotspot problem usually occurs sporadically, so it is difficult to obtain a large amount of affected data, in which could make the data unbalanced. In order to solve the above problems effectively, a prediction method of complaint hotspot problem based on users' signalling big data is proposed. In this method, the key fields which related to the user experience are firstly selected from the S1 interface. According to the characteristics of network business procedure, three types of features are extracted. The LightGBM classifier with different weight of each category is introduced to recognize the affected users. Experiments and the applications in network are carried out to validate the accuracy and efficiency of the proposal. 

The rest of the paper is organized as follows. In Section 2, we briefly discuss related work. Then we present our prediction method of complaint hotspot problem based on signalling data in Section 3. We evaluate our method in Section 4. Finally, we discuss future work and conclude in Section 5.
\par

\section{Related Work}
More recently, machine learning has been introduced for complaints handling and locating the failure of network. Chao et al. studied a self-updating network complaint hotspots location method based on kernel density estimation \cite{8094130}. The new hotspots were identified based on coordinates and the number of the customer complaints in the past few weeks. Then, the kernel function and the density function was introduced to sum the complaints scale of each location. Finally, the hotspots can be adjusted by the weight based on KPIs and signal quality. Tu\v{g}ba et al. proposed a new customer oriented smart diagnosis and solution model approach based on decision tree\cite{Kasapogullari2016Smart}. Zhang et al. introduced an active network optimization approach based on measurement report\cite{Kai2012WCDMA}. Jin et al. presented the potential network problems identification method based on customer tickets or LRT, a smartphone application for collecting large-scale feedback from mobile customers \cite{Jin2011Making, Jin2012Large}. Feyzullah et al. designed a LDA based topic model to analyze the customer complaint and showed top topic distributions\cite{kalyoncu2018customer}. Considering the inaccuracy of user complaints, Li Yan et al. created a rural distribution network fault location algorithm based on fault complaint information\cite{Yan2011Rural}. Ibrahim et al. studied a topic modeling approach for detecting network problems from customer complaints\cite{Yigit2017An}. Qing et al. combined the cellphone customer complaints data and equipment failures data to predict the equipment failures with the decision tree\cite{yang2017correlation}. The method based on KPIs and measurement reports can hardly locate the affected users in time, and take active care of the affected users. The method based on the user complaints cannot locate the affected area and troubleshoot potential problems in time. And this kind of method belongs to the way of post-complaint processing, which has a large delay. In addition, due to the inaccuracy of user complaints, the detection or prediction result may be inaccuracy.Jin et al. proposed NEVERMIND to predict customer tickets and locate problems of the DSL network \cite{jin2010nevermind}. Unlike mobile networks, the DSL network is stationary, so it can be seen as a failure detection model. Moreover, NEVERMIND processes weekly grained data, which cannot meet the real-time response requirements of mobile networks.

To solve the above problems, we proposed a prediction method of complaint hotspot problem based on users' signalling big data. Real-time monitoring of signalling data is used to instantly recognize whether complaint hotspot problem occurs, then locate the affected area and output the potential affected user group. The results of our method have high accuracy and interpretability. 

\section{PROPOSED METHOD}
Once the complaint hotspot problem occurs in the mobile network, the user experience will be affected more or less, which is usually reflected in the user's signalling data  and user plane data in real time. For example in the LTE network. The architecture of LTE network is shown in Figure 1. The signalling interaction and the user plane data between users and the network are reflected in S1 interface data. The S1-MME interface supports functionalities such as paging, handover, context management of the UE, E-RAB management, which reflects the experience of the establishment of a communication channel link between the user and the network. The S1-U between the eNodeB and the Serving GW is a user plane interface, carrying user data received from the terminal or the server of application, which reflects the user experience in the process of business transmission. This paper focuses on the prediction of LTE network complaint hotspots problem with the S1-MME and S1-U interface data.

\begin{figure}[tp]
	\centering
	\includegraphics[width=3.3in,height=1.6in]{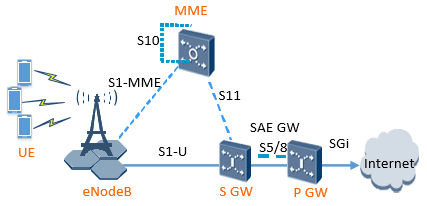}
	\setlength{\abovecaptionskip}{0.cm}
	\setlength{\belowcaptionskip}{-0.cm}
	\caption{The architecture of LTE network.}
\end{figure}

\subsection{Preprocessing}
We have selected 30 key features related to the user experience based on our professional knowledge of network capability and user service procedure, including 6 key features of S1-MME data and 24 key features of S1-U data.  There are 9 enumerated features, such as procedure type 1 for attach, 2 for service request, L4protocol 1 for TCP, and 2 for UDP. The other 21 features, such as upload traffic, download traffic, are numeric type features.

To clean up the S1-MME and S1-U data, consistency check is used to remove invalid data, erroneous data and duplicate data. Considering that there are missing values in the data, for numeric type data, the average value of this data can be utilized to fill in according to the logical relationship.

In order to characterize the user's business process in more detail, we concatenate of the control plane with sequential associations to form new features. For example, if the procedure type value `1' and a procedure status value `0' are concatenated into one feature, then the new value `10' means the user attach is successful. In the end a total of 33 additional features are obtained.

\subsection{Feature Extraction}

\subsubsection{One-hot feature extraction}
The value of the enumerated features is artificially defined without the meaning of size or sequence. Thus, they cannot be used as input to machine learning models directly, such as procedure status value '0' means success, value '1' means failure, value '255' means timeout. As with the processing in machine learning, one-hot encoding is required for these fields. The matrix processed by these fields is not so sparse, so the binary features converted from the categorical features can be used as the input of the model. Assuming that there are n kinds of valid values for a categorical feature, the feature can be encoded into a one-hot vector with n values, where only one value is 1 and others are 0, as shown in Figure 2.

\begin{figure}[hp]
	\centering
	\includegraphics[width=2.8in,height=1.77in]{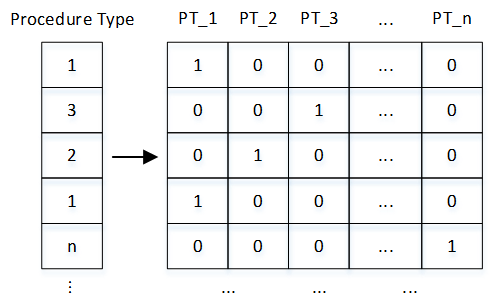}
	\setlength{\abovecaptionskip}{0.cm}
	\setlength{\belowcaptionskip}{-0.cm}
	\caption{One-hot feature extraction.}
\end{figure}

\subsubsection{Statistical feature extraction}
Each operation of the user generates a signal data, and analyzing each signalling data separately makes no sense since there are a large number of retrying and disaster recovery mechanisms in the communication network to ensure the user experience. For example, for a normal user, a signalling process (such as a service request) may fail due to a terminal or a special reason. But it will often succeed  after retrying 1 or 2 times, it will not affect the user experience. However, the users affected by complaint hotspot problem often have frequent failures even timeouts in multiple signalling, which affects the user experience.  Considering the particularity of the communication network, the statistical features of the user signalling data are extracted as the input of the model, instead of taking each XDR record as a sample. 

For the categorical features, we calculate the total number of services and the number of signal generated by a user for a period of time, such as counting the number of user service request failures within 5 minutes, which can be done by adding the one-hot vectors.

For the numeric features, the maximum, minimum, average, standard deviation, median and sum over a period of time are calculated to describe the user's experience in detail. For example, compared to unaffected users, the affected users may have a longer TCP link response delay. These six-dimensional statistical features can make a significant contribution to distinguishing between affected and unaffected users. And then the signalling data of control plane and user plane are combined together.

\subsubsection{Differential feature extraction}
In order to extract the temporal trend of user's data, we introduce the first-order difference and second-order difference characters. First-order difference equation is denoted as Equation(1).
\begin{equation}
	\Delta {y_t} = y\left( {t + 1} \right) - y\left( t \right),t = 0,1,2, \ldots ,n\
\end{equation}

\noindent where y denotes the processed signalling data of one user at time \emph{t}, and $\Delta {y_t}$ denotes the first-order difference sequence at this time.
\par
Considering that the first-order difference sequence still contains a long-term increasing trend, a difference operation is performed again to obtain a second-order difference sequence, denoted as Equation(2).

\begin{equation}
	\Delta \left( {\Delta {y_t}} \right) = \Delta {y_{t + 1}} - \Delta {y_t}\
\end{equation}

By performing two differentials on the data, some long-term trends implied in the original sequence can be extracted. After the feature extraction process above, all the features finally used for classification are obtained.

\subsubsection{Classifier}
Due to the particularity of the data, we faced two problems when choosing a classification model. The first problem is that the raw data contains a certain amount of noise, and we need to choose a classification model with strong generalization ability. The second one is that in order to do some follow-up analysis, we need to find the key component by sorting the importance of features. Considering the above two issues, GBDT(Gradient Boosting Decison Tree) is a smart choice, since GBDT has strong generalization ability and interpretability\cite{friedman2001greedy, friedman2002stochastic}.


Considering that GBDT is used as a regression model, a new decision tree will be built in each round. When building the \emph{t}-th decision tree in \emph{t}-th round, GBDT uses the residual of the regression values of the previous \emph{t}-1 decision trees on all samples as the value to be fitted. When GBDT is used for classification, the model will build \emph{K} decision trees at the same time in each round, corresponding to \emph{K} classifications, instead of building only one decision tree in each round. In the \emph{t}-th round, the softmax value of the regression value   $f_{m}^{t-1}$ of the \emph{m}-th decision tree generated from the previous \emph{t}-1 round of a sample $x_{i}$ indicates the probability that the sample belongs to the classification \emph{m}, as shown in Equation(3):

\begin{equation}
	P_{m}^{t}=\frac{e^{f_{m}^{t}(x_{i})}}{\sum_{p=1}^{K}e^{f_{p}^{t}(x_{i})}}
\end{equation}

Let the true label of the sample xi be represented as a one-hot type vector $\left [ q_{m,i}\mid m= 1,2,\cdots,K \right ]$. Only if the sample $x_{i}$ belong to the class \emph{k}, the $q_{k,i}$ equals 1. Then residual error of the \emph{m}-th classifier of $x_{i}$ in \emph{t} round is $\sum\limits_{i = 1}^N {{q_{m,i}} - P_m^t({x_i})}$.

Although GBDT performs well on the general data set, it still has two problems. One problem is that  when the data set is large or the feature dimension is high, the efficiency and scalability of the algorithm are difficult to meet the demand. The other problem is that the model does not work well on unbalanced data set. Aiming to solve these problems, we introduced a model named LightGBM\cite{ke2017lightgbm}, which is an efficient GBDT model.

For the first problem, LightGBM is optimized in two aspects. On the one hand, the model uses the GOSS algorithm to optimize the sampling process during training. On the other hand, the EFB algorithm is used to reduce the sample feature dimension. These two optimization methods help the model reduce the calculated amount and improve the performance of the algorithm.

The model improved by GOSS does not traverse all samples to find the optimal threshold according to the principle of maximum information gain when performing node splitting on a feature \emph{j}. Instead, the model first sorts the samples according to the gradient values, then selects $a\%$ samples with the largest gradient value, and randomly selects \emph{b} samples from the remaining $\left ( 1-a \right )\%$ of the samples to form a new training set to build the classifier. The \emph{b} samples with smaller gradients are multiplied by a larger coefficient. The value of information gain calculation when splitting a node can be expressed as Equation(4):
\begin{equation}
	\begin{split}
		\widetilde{V}_{j}\left ( d \right )=\frac{1}{n}\left ( \frac{\left ( \sum _{x_{i}\in A_{l}}g_{i}+\frac{1-a}{b}\sum _{x_{i}\in B_{l}}g_{i} \right )}{n_{l}^{j}\left ( d \right )} \right )\\
		+\frac{1}{n}\left ( \frac{\left ( \sum _{x_{i}\in A_{r}}g_{i}+\frac{1-a}{b}\sum _{x_{i}\in B_{r}}g_{i} \right )}{n_{r}^{j}\left ( d \right )} \right )
	\end{split}
\end{equation}

\noindent where  $A_{l}=\left \{ x_{i}\in A:x_{ij}\leq d \right \}$, $A_{r}=\left \{ x_{i}\in A:x_{ij}> d \right \}$, $B_{l}=\left \{ x_{i}\in B:x_{ij}\leq d \right \}$, $B_{r}=\left \{ x_{i}\in B:x_{ij}> d \right \}$, \emph{d} is the optimal threshold culculated by the model.
\par
For sparse high-dimensional data, LightGBM can bundle exclusive features through the EFB algorithm to achieve the target of reducing feature numbers. That means the complexity of the histogram creation will be reduced from \emph{O}(data * feature) to \emph{O}(data * bundle), when the histogram is created, thus accelerating the training process of LightGBM.

This approach allows LightGBM to use non-full data for training each time when fitting a new tree, which reduces computational load from the training data dimension, increases decision tree diversity, and improves generalization capability.

To solve the second problem, LightGBM can make some adjustments for weights of positive samples.

The flow chart of the proposed method is shown in Figure 3.

\begin{figure}[hp]
	\setlength{\abovecaptionskip}{0.cm}
	\setlength{\belowcaptionskip}{-0.cm}
	\centering
	\includegraphics[width=3.3in,height=4.58in]{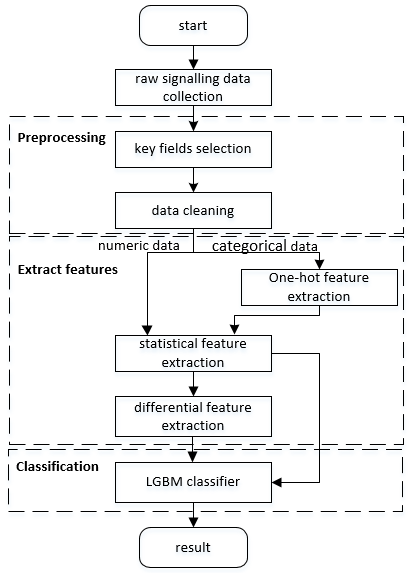}
	\caption{The general block diagram of proposed method}
\end{figure}
\par
\section{EXPERIMENT AND APPLICATIONS}
To validate the effectiveness of the proposed method, experiments are conducted based on atmospheric waveguide case. Atmospheric waveguide phenomenon causes the signal to cross the protection time slot and seriously interfere with the uplink of the remote base station. That phenomenon leads to deterioration of VoLTE voice call quality and slow speed of LTE Internet access, thereby affecting the user experience and causing bulk complaints.

In order to ensure the robustness of the model and characterize the affected user data in more detail and comprehensively, three types of user data are selected to construct a negative sample set, which were all labeled by the field technicians, including the unaffected users during the atmospheric waveguide period, other complaint users, and normal users. Thus, we collected 702598 samples of S1-MME interface data and 2073038 samples of S1-U interface data. 30 key fields related to user experience are selected according to the professional knowledge of network and user service procedure, such as procedure type, procedure status, request cause, failure case, app type code, app type whole, L4 protocol, upload/download traffic. Then the invalid data, erroneous data and duplicate data are removed by consistency check. The average of the numeric field is used to fill the missing values. There is sequential logical relationship between the fields, so these pairs of fields are spliced to form new fields, such as procedure type - procedure status, procedure type - request cause, procedure type - failure cause.

The categorical fields are converted to binary features by one-hot. In order to extract the statistical feature, the data is sliced by time. The choice of duration is based on the following considerations. In practice, atmospheric waveguide phenomenon usually affects users' service during a few minutes. If the slicing period is too long, the statistical features may be so smooth that it is same as normal users'. If the slicing time is too short, the statistical features may not effectively distinguish the complaint hotspot problem from the sporadic small-scale network failure. Thus, the 5 minutes is finally chosen. After sliced, the statistical features of each slice are extracted. Then, the S1-MME data and the S1-U data are spliced in time. Figure 4. shows the statistical characteristics of TCP construct link ack time. It can be seen that the statistical characteristics of the affected users and the normal users are different.

\begin{figure}[hp]
	\setlength{\abovecaptionskip}{0.cm}
	\setlength{\belowcaptionskip}{-0.cm}
	\centering
	\includegraphics[width=3.3in,height=2.9in]{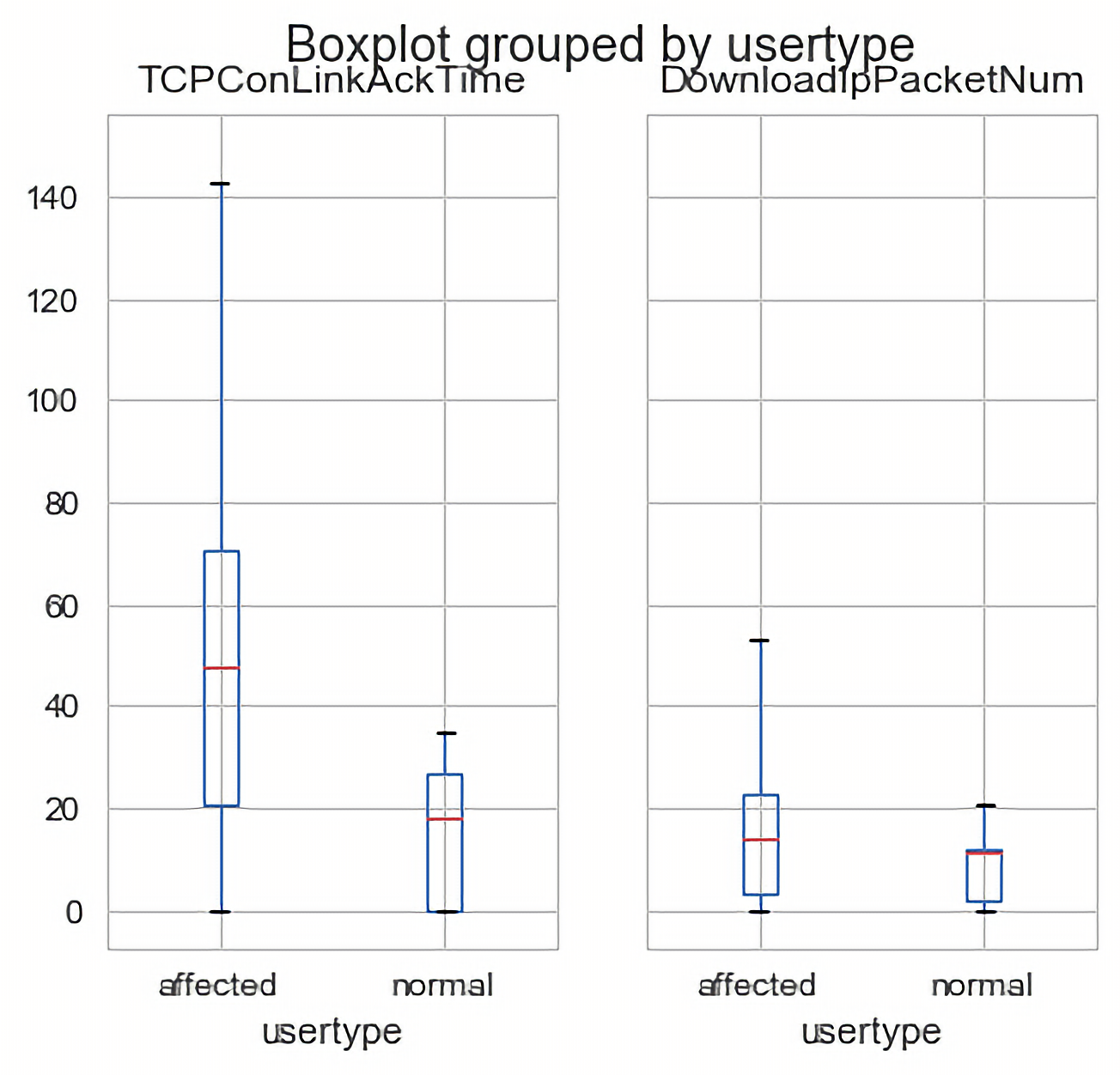}
	\caption{Statistical feature. The box diagram of the TCP construct link ack time and download IP packet number of affected users and normal users.}
\end{figure}
\vspace{0cm}
Through above steps, 626 statistical features are extracted. Subsequently, the difference features are extracted. Figure 5. shows the data trend after the first-order difference and the second-order difference.

\begin{figure}[htp]
	\setlength{\abovecaptionskip}{0.cm}
	\setlength{\belowcaptionskip}{-0.cm}
	\centering
	\includegraphics[width=3.3in,height=2in]{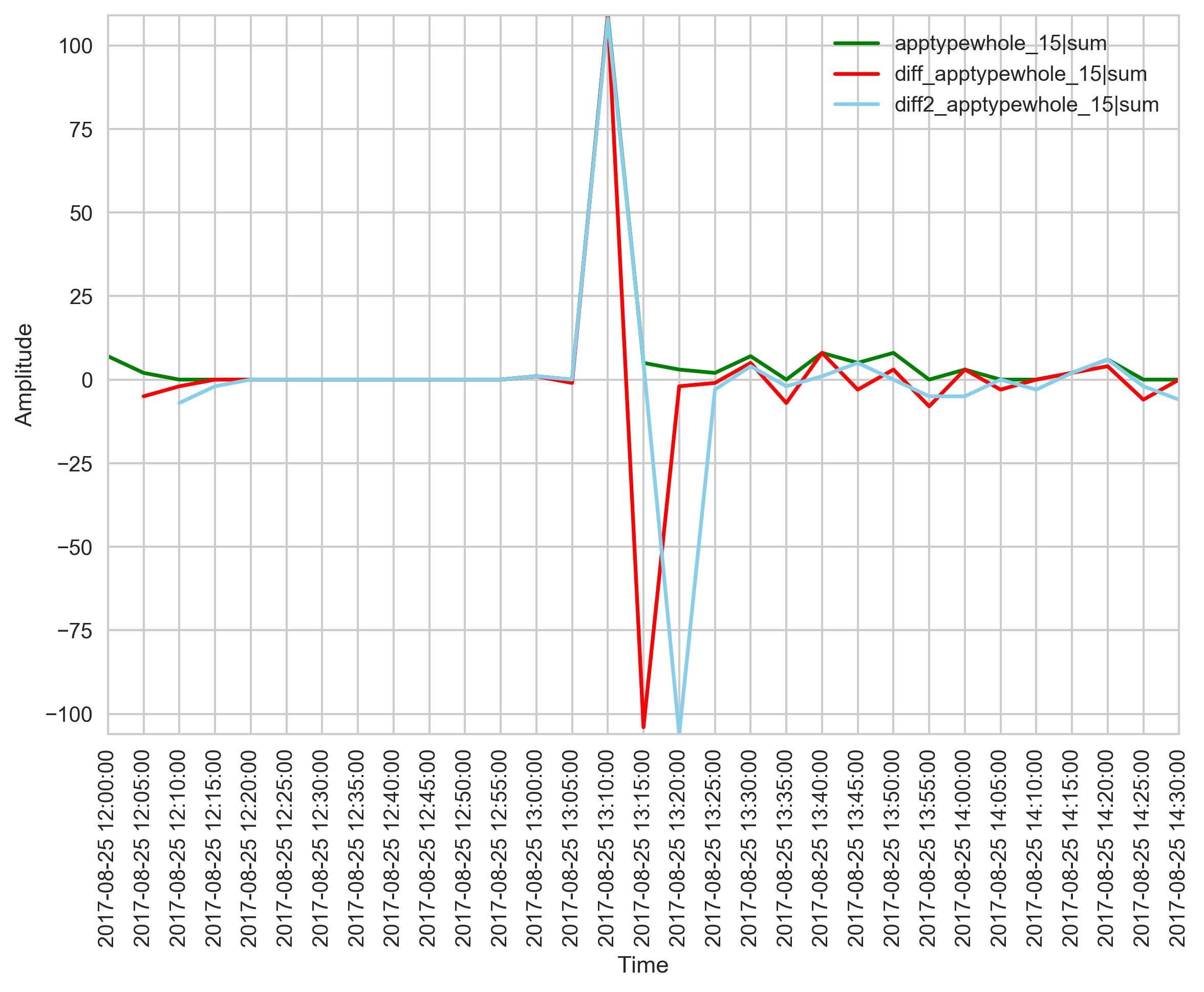}
	\caption{Differential feature. The picture shows the first-order difference and the second-order difference of the feature, which denotes the total number of apptypewhole value 15 in five minutes.}
\end{figure}
\vspace{0cm}
Finally, the 252325 by 1878 sample matrix, including less than 10$\%$ of negative samples, is divided into train set and test set by 7:3. Considering the imbalance of the sample, the precision and recall cannot fully reflect the classification performance. Therefore, we use the F1 score that takes both into account to evaluate the results. While setting the largest number of leaves as 120, the LightGBM model converged after 230 iterations. The relationship between F1score and weights of positive samples is shown in Figure 6. From this relationship, it can be concluded that the performance is best when the weight of positive samples is 5.The ROC curve and PR curve at this time are shown in Figure 7 and Figure 8.
\begin{figure}[hp]
	\centering
	\setlength{\abovecaptionskip}{0.cm}
	\setlength{\belowcaptionskip}{-0.cm}
	\includegraphics[width=3.3in,height=2in]{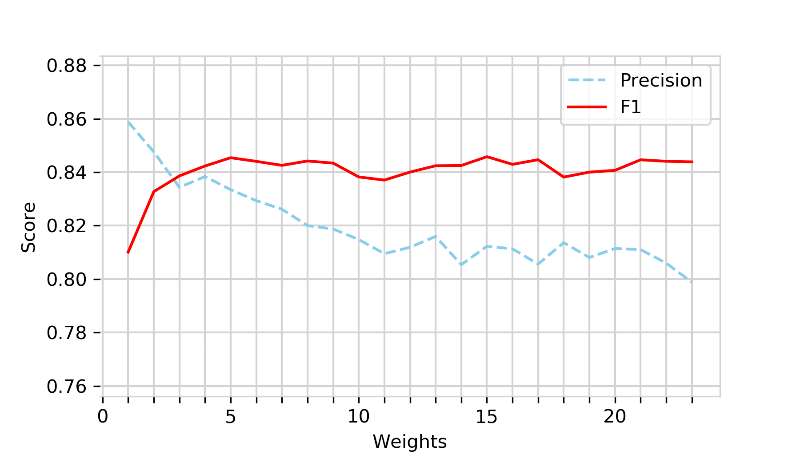}
	\setlength{\abovecaptionskip}{0.cm}
	\setlength{\belowcaptionskip}{-0.cm}
	\caption{The relationship between F1 score, precision and weights.}
\end{figure}
\vspace{0cm}
\begin{figure}[tp]
	\centering
	\setlength{\abovecaptionskip}{0.cm}
	\setlength{\belowcaptionskip}{-0.cm}
	\includegraphics[trim=0mm 0mm 0mm 5mm,clip,width=3.3in,height=2in]{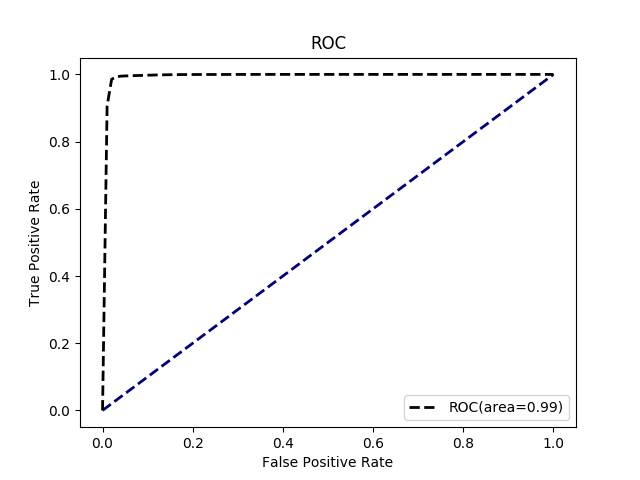}
	\setlength{\abovecaptionskip}{0.cm}
	\setlength{\belowcaptionskip}{-0.cm}
	\caption{The AUC Evaluation. Diagram shows the ROC curve based on the LightGBM.}
\end{figure}
\vspace{0cm}
\begin{figure}[tp]
	\centering
	\setlength{\abovecaptionskip}{0.cm}
	\setlength{\belowcaptionskip}{-0.cm}
	\includegraphics[width=3.3in,height=2.48in]{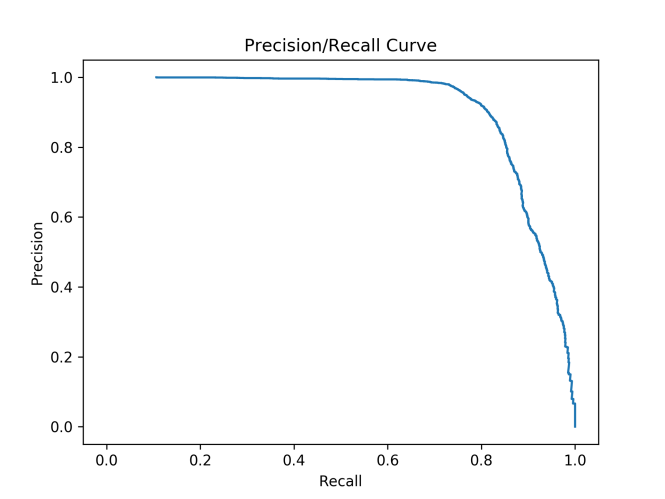}
	\setlength{\abovecaptionskip}{0.cm}
	\setlength{\belowcaptionskip}{-0.cm}
	\caption{The PR Curve.}
\end{figure}

Through the feature importance calculation, we get the order of importance of all features. The five most important features of the user plane and the control plane are listed in Table 1. Furthermore, we analyze the distribution of these features in both positive and negative samples. Taking 'Number of signalling data' as an example, Figure 9 shows the distribution difference of the feature in positive and negative samples. Through the analysis based on network business logic, we find that affected users have no signalling data of control plane in most of slicing period due to the inability to perform normal business processes. 
\begin{table}[tp]
	\centering
	\begin{tabular}{p{3.6cm}|p{3.6cm}}
		\toprule
		\textbf{Control plane} & \textbf{User plane} \\
		\midrule
		Number of signalling data & Maximum of the spendtime \\
		\hline Sum of paging success number & Standard variance of windowSize  \\
		\hline Total number of E-RAB release & Minimum of TCPSynNum \\
		\hline Total number of signalling data & Minimum of upload traffic \\
		\hline Total number of authentication failure & Sum of the spendtime \\
		\bottomrule
	\end{tabular}%
	\setlength{\abovecaptionskip}{0.cm}
	\setlength{\belowcaptionskip}{-0.cm}
	\caption{Features of TOP importance.}
	\label{tab:addlabel}%
\end{table}%

\begin{figure}[tp]
	\centering
	\setlength{\abovecaptionskip}{0.cm}
	\setlength{\belowcaptionskip}{-0.cm}
	\includegraphics[width=2.8in,height=1.78in]{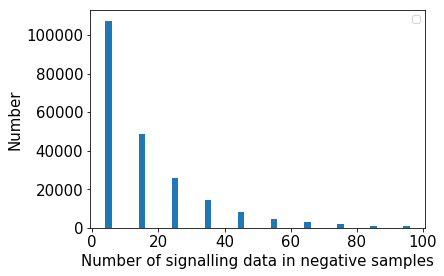}
\end{figure}
\begin{figure}[tp]
	\centering
	\setlength{\abovecaptionskip}{0.cm}
	\setlength{\belowcaptionskip}{-0.cm}
	\includegraphics[width=2.8in,height=1.78in]{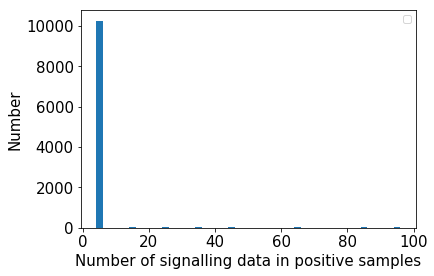}
	\caption{The distribution of the 'Number of signalling data' in positive and negative samples.}
\end{figure}
Besides, the comparative experiment is conducted. The result is shown in Table 2. In LightGBM training process we increase the weight of the negative samples. The result shows the LightGBM model can recognize more positive samples than others. Due to the actual service needs in the network and its efficiency, accuracy, interpretability, LightGBM is more suitable for the prediction of the complaint hotspot problem.
\begin{table}[tp]
	\centering
	\begin{tabular}{c|c|c}
		\toprule
		\textbf{Method} & {\textbf{Precision}} & {\textbf{F1 score}} \\
		\midrule
		\textbf{LightGBM} & 83.35\% & \textbf{84.54\%} \\
		Decision Tree & 75.71\% & 71.06\% \\
		Random Forest & 83.25\% & 78.12\% \\
		GBDT  & 85.35\% & 80.90\% \\
		DNN   & 75.59\% & 78.81\% \\
		CNN   & 79.25\% & 76.83\% \\
		\bottomrule
	\end{tabular}%
	\setlength{\abovecaptionskip}{0.cm}
	\setlength{\belowcaptionskip}{-0.cm}
	\caption{The comparison of LightGBM classifier with Decision Tree, Random Forest, GBDT, DNN and CNN.}
\end{table}%

At present, the proposed method has been deployed in the network of a certain province in China. By introducing the prediction model into the complaint pre-processing section, the pre-processing of the complaint problem can be implemented. The prediction time can be compressed to within 5 minutes, and the accuracy rate is 87$\%$ with 17.1$\%$ decline in complaints rate.

\section{CONCLUSIONS}
This paper presents a prediction method of the complaint hotspot problem based on signalling big data. By preprocessing the original S1 interface signalling data, the key fields are selected and cleaned. One-hot features, statistical features and differential features are extracted to characterize the differences between the affected users and unaffected users. The LightGBM classifier which has strong generalization ability, interpretability and the adjustable weight is used to recognize the affected users. Through experiments on atmospheric waveguide case, it is proved that the proposed method can be used to detect the complaint hotspot problem more proactively with better performance, and it is more suitable for the prediction of the complaint hotspot problem. The applications used in the commercial network in a province of China proves the feasibility and efficiency of the proposed method. The proposed method can help the operators locate the problems, output the potential affected area and user groups in the first time, then take customer care before user complaints, thereby avoiding bulk complaints.

\renewcommand*{\bibfont}{\large}

\bibliographystyle{acm_unsort}
\large
\bibliography{sigproc}

\end{document}